\documentclass[12pt]{article}
\newcommand{\fighere}[1]{#1} 
\newcommand{\edo}{\end{document}} 

\usepackage{epsf,latexsym,amssymb}

\usepackage{graphicx}

\usepackage{amsmath}  

\newcommand{\R}{{\mathbb R}}  

\newcommand{\abs}[1]{\left\vert #1 \right\vert}

\newtheorem{theorem}{Theorem}
\newtheorem{itlemma}{Lemma}[section] 
\newtheorem{itproposition}[itlemma]{Proposition}
\newtheorem{itcorollary}[itlemma]{Corollary}

\newenvironment{lemma}{\begin{itlemma}\rm}{\end{itlemma}} 
\newenvironment{corollary}{\begin{itcorollary}\rm}{\end{itcorollary}}
\newenvironment{proposition}{\begin{itproposition}\rm}{\end{itproposition}}

\newcommand{\be}[1]{\begin{equation}\label{#1}}
\newcommand{\ee}{\end{equation}}
\newcommand{\bl}[1]{\begin{lemma}\label{#1}}
\newcommand{\bt}[1]{\begin{theorem}\label{#1}}
\newcommand{\bp}[1]{\begin{proposition}\label{#1}}
\newcommand{\bc}[1]{\begin{corollary}\label{#1}}
\newcommand{\el}{\mybox\end{lemma}}
\newcommand{\et}{\qed\end{theorem}}
\newcommand{\ed}{\mybox\end{definition}}
\newcommand{\ep}{\mybox\end{proposition}}
\newcommand{\epr}{\end{proof}}
\newcommand{\bpr}{\begin{proof}}

\newcommand{\ecs}{\end{corollary}}
\newcommand{\els}{\end{lemma}}
\newcommand{\eps}{\end{proposition}}
\newcommand{\halmos}{\rule{1ex}{1.4ex}}
\newcommand{\qed}{\hfill \halmos} 
\newcommand{\mybox}{\hfill $\Box$} 
\newcommand{\beq}{\begin{eqnarray}}
\newcommand{\eeq}{\end{eqnarray}}
\newcommand{\beqn}{\begin{eqnarray*}}
\newcommand{\eeqn}{\end{eqnarray*}}
\newcommand{\bi}{\begin{itemize}}
\newcommand{\ei}{\end{itemize}}

\newenvironment{proof}{\noindent {\em Proof}.\ }{\hspace*{\fill}$\halmos$\medskip}

\newcommand{\Q}{\Theta}

\newcommand{\A}{P}
\newcommand{\B}{Q}

\title{Remarks on Feedforward Circuits, Adaptation,\\ and Pulse Memory}
\author{Eduardo D.\ Sontag\\
Department of Mathematics\\
Rutgers University\\
New Brunswick, NJ 08903}

\begin{document}

\maketitle

\subsubsection*{Abstract}

This note studies feedforward circuits as models for perfect adaptation to
step signals in biological systems.  A global convergence theorem is proved in
a general framework, which includes examples from the literature as particular
cases.  A notable aspect of these circuits is that they do not adapt to pulse
signals, because they display a memory phenomenon.  Estimates are given of the
magnitude of this effect.

\section{Introduction}

Feedforward circuits have been often proposed for adaptation to constant
signals in biological systems.  Indeed, the review paper~\cite{sniffers}
gives a chemical reaction model, called there a ``sniffer'' and shown in
Figure~\ref{sniffer_reactions}, as the paradigm for perfect adaptation.
\fighere{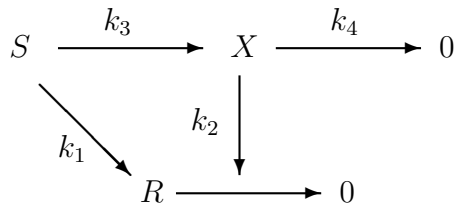
\begin{figure}[h,t]
\begin{center}
\setlength{\unitlength}{3000sp}
\begin{picture}(3542,1560)(2101,-3736)
\thicklines
\put(2401,-2461){\vector( 1, 0){1200}}
\put(4201,-2461){\vector( 1, 0){1200}}
\put(3376,-3661){\vector( 1, 0){1200}}
\put(2251,-2761){\vector( 1,-1){750}}
\put(3901,-2686){\vector( 0,-1){825}}
\put(4726,-3736){$0$}
\put(3076,-3736){$R$}
\put(2000,-2536){$S$}
\put(3826,-2536){$X$}
\put(5551,-2536){$0$}
\put(4651,-2311){$k_4$}
\put(3500,-3136){$k_2$}
\put(2401,-3361){$k_1$}
\put(2776,-2311){$k_3$}
\end{picture}
\caption{``Sniffer'' network, from ~\protect{\cite{sniffers}}}
\label{sniffer_reactions}
\end{center}
\end{figure}}
The chemical species $S$ acts as a ``signal'', and the species $R$ is viewed as
a ``response'' element.  The third species, $X$, is an intermediate species.
The species $S$ directly helps promote the formation of $R$ 
(arrow labeled ``$k_1$''), and also the formation of $X$
(arrow labeled ``$k_3$'').
On the other hand, $X$ also enhances degradation of the species $R$ (vertical
arrow labeled ``$k_2$''), and thus $S$ also acts through $X$ as an 
inhibitor of $R$, counteracting the positive direct effect.
This ``incoherent'' counterbalance between a positive and a negative effect
gives rise to a regulation property.

Mathematically, the model is described in~\cite{sniffers} as a system of
two coupled differential equations for the concentrations of the substances in
question, using mass-action kinetics:
\begin{subequations}
\label{sniffer_equations1}
\begin{eqnarray}
\dot x &=& k_3s - k_4x\\
\dot r &=& k_1s - k_2xr
\end{eqnarray}
\end{subequations}
where dot is used to indicate time derivative.
The key fact is that in steady state and for nonzero constant signals $S$,
the concentration of $R$ equals $\frac{k_1k_4}{k_2k_3}$, and this value is
independent of the actual value of $S$.  (This follows simply by setting the
right-hand sides of the equations to zero and solving for $r$.)

Similar constructions have been given in other biological investigations of
adaptation, notably in~\cite{iglesias05}, where various possible chemical
networks are proposed for modeling adaptation by the chemotaxis pathway of
\emph{Dictyostelium}.

A common feature of these models is that they have the form of a stable
linear system which in turn drives a one-dimensional (generally nonlinear)
system, whose
state-variable represents the response that should adapt.
The interconnection is set up so that the whole system becomes a ``feedforward
circuit''~\cite{alon}.

\subsubsection*{General framework}

The strength of the external input signal (a non-negative real number) will be
denoted by
$u$, the state of the linear system (an $n$-dimensional vector) by $x$, and
the state of the driven response (a real number) by $y$. 
Then, a mathematical description of the evolution of concentrations of the
various signals is given by a system of $n+1$ differential equations as follows:
\begin{align}
\label{system}
\begin{split}
\dot x &= Ax + bu\\
\dot y &= c(y)x + d(y)u
\end{split}
\end{align}
where $A$ and $b$ are a constant stable matrix and column vector
respectively, $A\in \R^{n\times n}$, $b\in \R^{n\times 1}$, and $c(y)$ and $d(y)$ are
continuous functions of $y$ so that, 
for each $y$, $c(y)\in \R^{1\times n}$ and $d(y)\in \R$.
System~(\ref{system}) is a system with inputs and outputs, in
the standard sense of control theory~\cite{mct}.
The state variables $x(t)$ and $y(t)$ take values in some subsets ${\cal X}\subseteq \R^n$
and ${\cal Y}\subseteq \R$ respectively, where ${\cal Y}$ is 
a closed, possibly unbounded,
interval.
The sets ${\cal X}$ and ${\cal Y}$ can be used in order to impose non-negativity and/or
mass conservation constraints.
Enough regularity is assumed so that, for every
non-negative constant input $u_0$, and every initial condition
$(x_0,y_0)\in {\cal X}\times {\cal Y}$, the equations~(\ref{system}) have a unique solution
$(x(t),y(t))\in {\cal X}\times {\cal Y}$ defined for all $t\geq 0$.

For example, in the ``sniffer'' reactions 
(\ref{sniffer_equations1})
from~\cite{sniffers}, and writing
$u$, $x$, and $y$ instead of $s$, $x$, and $r$ respectively, one has
${\cal X}={\cal Y}=\R_{\geq 0}$, $A=-k_4$, $b=k_3$, $c(y)=-k_2y$, and $d(y)=k_1$ (constant).
In these notations Equations~(\ref{sniffer_equations1}) become:
\begin{subequations}
\label{sniffer_equations}
\begin{eqnarray}
\dot x &=& - k_4x + k_3u \\
\dot y &=&  - k_2xy + k_1u.
\end{eqnarray}
\end{subequations}

\subsubsection*{Adaptation to step inputs}

For each \emph{nonzero} constant input $u_0$, the steady states $(x_0,y_0)$ of
the system~(\ref{system}) are obtained by first setting $\dot x=0$, which gives
$x_0=-A^{-1}bu_0$ 
(the inverse is well-defined because $A$ was assumed to be stable,
and in particular all its eigenvalues are nonzero), and then substituting into
the left-hand side of the $y$-equation.  There obtains the following algebraic
equation: 
\be{eq:alg}
d(y) - c(y)A^{-1}b \;=\; 0
\ee
(after canceling out $u_0\not= 0$).
A key hypothesis made from now on (and which is satisfied in all
the cited examples) is that \emph{there is a unique solution} $y=y_0$ of the
algebraic equation~(\ref{eq:alg}), and that this solution is an asymptotically
stable state for the reduced system
\[
\dot y = (d(y) - c(y)A^{-1}b)u_0
\]
that would
result if $x(t)$ were already at its steady state $-A^{-1}bu_0$.
To be precise, the following hypothesis will be imposed:
$$
(\exists \, y_0\in {\cal Y}) \;
(\forall \, y\in {\cal Y}) \;
\left[\,\left(y-y_0\right)\,\left( d(y) - c(y)A^{-1}b\right) \,<0
 \, \right]
\eqno\mbox{(\bf H)}
$$
(that is, $\dot y = (d(y) - c(y)A^{-1}b)u_0$ is positive when $y<y_0$ and negative
when $y>y_0$).
It is fundamental to observe that $y_0$ (though not, of course, $x_0$) is
independent of the  particular numerical value of $u_0$.

Proposition~\ref{theorem-adapt} shows that, assuming
boundedness of trajectories,
systems~(\ref{system}) ``adapt'' to nonzero constant
signals $u_0$ (``step signals''), in the sense that all the solutions of
the system~(\ref{system}) converge to the above steady state $(x_0,y_0)$,
where $y_0$ is independent of $u_0$.

Take again as an example the equations (\ref{sniffer_equations})
from~\cite{sniffers}, now with all $k_i=1$,
Figure~\ref{fig:4inputs}(b) plots the response to the piecewise constant
input with nonzero values that is shown in Figure~\ref{fig:4inputs}(a).
It is clear that this response adapts to the value $y_0=1$.
\fighere{\begin{figure}[h,t]
\begin{center}
\includegraphics[scale=1]{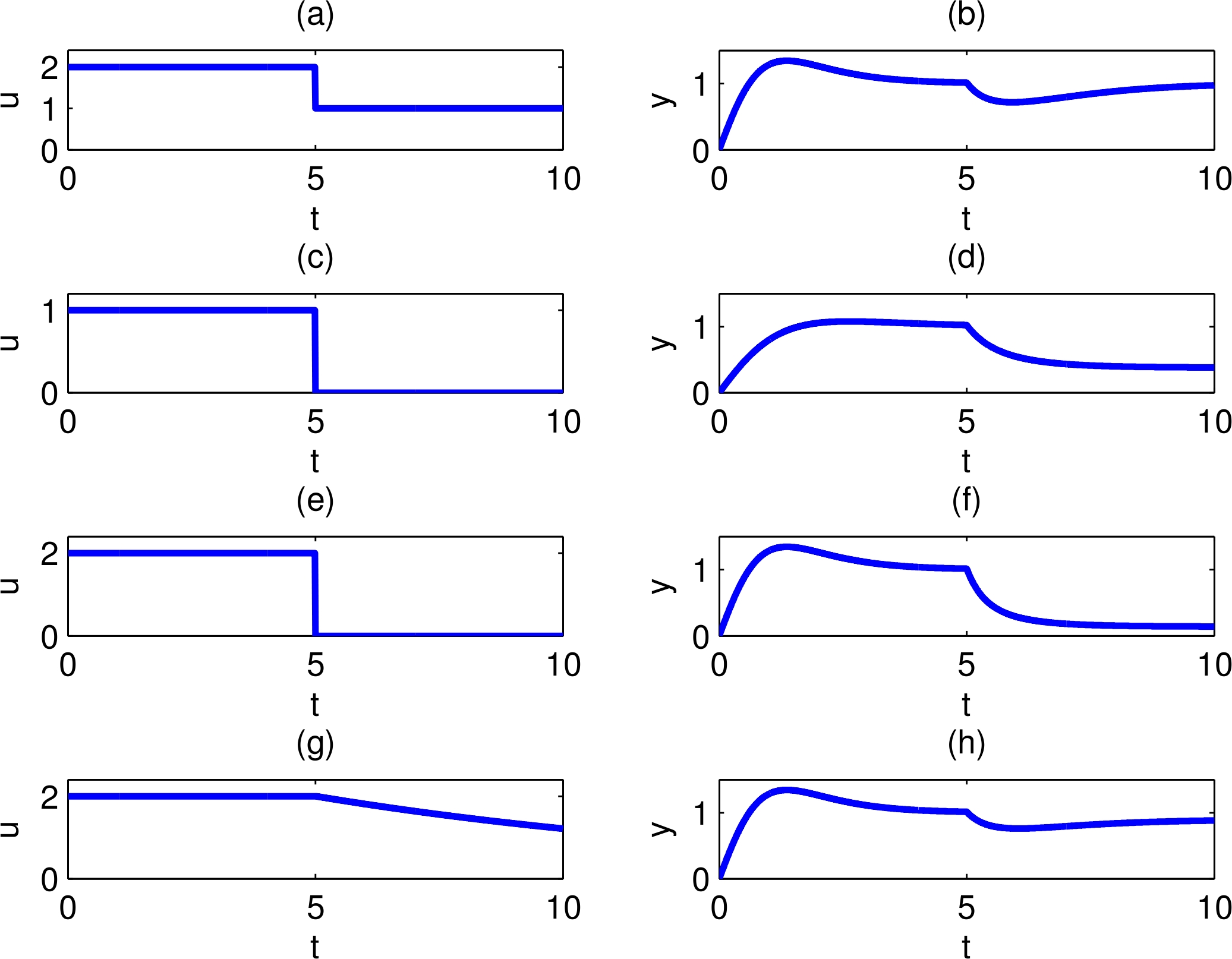}
\label{fig:4inputs}
\caption{%
Plots for example discussed in text.  In left column are inputs $u(t)$,
in right column are responses $y(t)$.  Initial conditions are $x(0)=y(0)=0$.
$u(t)$ is constant up to time $t=5$, switches to another constant value
at time $t=5$.
(a,b) $u$ switches to a nonzero value, $y$ adapts again.
(c,d) Pulse of magnitude $1$: asymptotic value of $y$ is 
$\approx 1/e \approx 0.37$ after pulse.
(e,f) Pulse of larger magnitude $2$: asymptotic value of $y$ is 
smaller: $\approx 1/e^2 \approx 0.14$ after pulse.
(g,h) Exponential decaying $u(t)=2e^{(t-5)/10}$ after time $t=5$
results in return to close to adapted value $y(\infty )\approx 0.9$
($u(t)$ plotted only on interval $[0,10]$ for ease of comparison).}
\end{center}
\end{figure}}

\subsubsection*{Memory of pulse inputs}

One of the main objectives of this note is to bring attention to
the following additional facts.  When $u_0=0$, that is, in 
the absence of an external signal, steady states are no longer unique.
Indeed, any vector of the form $(0,y)$ is a steady state.  This has
an important consequence for the behavior of system~(\ref{system}) when a
\emph{pulse} input is used.  A pulse is defined here as an input $u$ which has
the following form: $u(t)=u_0\not= 0$ for some interval $t\in [0,T]$, and $u(t)=0$
for $t>T$. 
Suppose that the interval is long enough ($T\gg1$) so that one may assume that
$x(T)$ and $y(T)$ are (approximately) in steady state: $x(T)\approx-A^{-1}bu_0$
and $y(T)\approx y_0$.
Upon removal of the external excitation at time $T$, the equations for the
system become $\dot x=Ax$ and $\dot y=c(y)x$ for $t>T$, with initial conditions
$x(T)$ and $y(T)$, so $x(t) \approx -e^{(t-T)A}A^{-1}bu_0$ for $t\geq T$. 
The solution of $\dot y = c(y)x$ starts at $y_0$ but adds a quantity which
integrates the effect of the nonzero function $x(t)$.
The response may then settle to a new value which is different than the
adapted value $y_0$.

Thus, feedforward systems for adaptation as those discussed here exhibit a
``memory'' effect with respect to (ideal) pulses.  This phenomenon
was apparently not remarked upon earlier. The present note discusses the
effect and provides estimates.

To illustrate the phenomenon with the simplest possible (if not biologically
meaningful) example, take $n=1$, $A=-1$, $b=1$, $c(y)=-1$, and $d(y)=1$.
One has that  $\dot y \approx -e^{T-t}u_0$, so $y(t)\rightarrow y_0-u_0<y_0$ as $t\rightarrow \infty $
(assuming that $u_0>0$).
In fact, the larger the $u_0$, the smaller the new asymptotic value
$y_0-u_0$ is.
For the nonlinear equations (\ref{sniffer_equations})
from~\cite{sniffers}, the same phenomenon holds.
Taking all $k_i=1$, 
Figure~\ref{fig:4inputs}(d) plots the response to a pulse of unit amplitude
and
Figure~\ref{fig:4inputs}(f) plots the response to a pulse of twice the
amplitude.  Once again, the response settles to a value that is smaller
when the amplitude of the pulse was larger.
The sharp cut-off of an ideal pulse plays an important role on this ``memory''
effect: when the input instead returns slowly enough to its baseline value,
an almost-adapted response is recovered, as shown in
Figure~\ref{fig:4inputs}(h).

\subsubsection*{Feedforward motifs in systems biology}

Feedforward circuits are ubiquitous in biology, as emphasized in~\cite{alon},
where (incoherent) feedforward circuits were shown to
be over-represented in \emph{E.coli} gene transcription networks
compared to other ``motifs'' involving three nodes.
Similar conclusions apply to certain control mechanisms in
mammalian cells~\cite{maayan_science05}.
A large number of papers have been devoted to the signal-processing
capabilities of the feedforward motif, notably~\cite{mangan06} which looked into
its properties as a ``change detector'' (essentially, sensitivity to changes
in the magnitude of the input signal), and~\cite{cournac} which studies its
optimality with respect to periodic inputs.
Comparisons with other ``three node'' architectures with respect to
the trade-off of sensitivity versus noise filtering are given in~\cite{hornung}.
Other references on feedforward circuits include~\cite{semsey}
(showing their over-representation at the interface of genetic and metabolic
networks),
\cite{wall} (classification of different subtypes of such circuits),
and~\cite{kim} (classification into
``time-dependent'' versus ``dose-dependent'' biphasic responses, which are
in a sense the opposite of adaptated responses).
The latter reference provides a large number of additional
incoherent feedforward input-to-response circuits, including:
EGF to ERK activation
\cite{kim16,kim18},
glucose to insulin release
\cite{kim28,kim30},
ATP to intracellular calcium release
\cite{kim33,kim32},
nitric oxide to NF-$\kappa$B activation
\cite{kim34},
microRNA regulation
\cite{kim12},
and many others.
Dealing specifically with adaptation properties of feedforward circuits,
and in addition to the papers~\cite{sniffers,iglesias05}, are the
paper~\cite{xu_mRNAfeedforward09} on microRNA-mediated loops,
and~\cite{Kremling}, which deals with the role of feedforward structures in
the robust behavior in \emph{E.coli} carbohydrate uptake via the carbohydrate
phosphotransferase system (an analogous metabolic mechanism is also discussed
in~\cite{voit-intricate}).

\subsubsection*{Outline of paper}

Section~\ref{sec:statements} has statements of the convergence results.
Section~\ref{sec:approximate} has a brief discussion of ``approximate''
adaptation by feedforward circuits.
Section~\ref{sec:memory} shows estimates of the
magnitude of the pulse memory effect.
Section~\ref{sec:proofs} has the proofs of the convergence results.
Section~\ref{sec:examples} revisits the motivating examples
and also briefly discusses the systems in~\cite{Kremling,xu_mRNAfeedforward09}.
Finally, it is known that, under appropriate technical assumptions,
perfect adaptation implies that the system may be written, after a
suitable nonlinear change of coordinates, as a system in which the
integral of the regulated quantify is fed-back, see for
instance~\cite{francis,hepburn,doyleIMP,imp03}.
This fact is not incompatible with the system
being a feedforward system, as remarked in Section~\ref{sec:imp}.
Section~\ref{sec:discussion} summarizes the results and speculates on
their significance.

\section{Statements of convergence results}
\label{sec:statements}

The main convergence result is as follows.
Note that, since 
the $x$-coordinate of a solution $(x(t),y(t))$ for constant $u$ always
converges (because of the stability assumption on the linear system) and hence
is bounded, asking that $(x(t),y(t))$ is bounded is the same as asking that
$y(t)$ is.

\bp{theorem-adapt}
Suppose that Property {\bf (H)} holds.
Then, for each step input $u\equiv u_0\not= 0$, and every initial condition, if 
the corresponding solution $(x(t),y(t))$ of~(\ref{system}) is
bounded,
then it converges to $(x_0,y_0)$.
\eps

Boundedness
is automatically satisfied if ${\cal Y}$ is itself a 
bounded
interval,
as is the case if mass conservation laws constrain the system dynamics.  More
generally, the following condition can be helpful.
It strengthens Property Property {\bf (H)} for small and for large values of
$y$:
\begin{align*}
&\!\!\!\!\!\!(\forall \, {u_0}>0)
(\exists \, {\bar \varepsilon >0})
\notag\\
&(\exists \, y_2\in {\cal Y}) \;
(\forall \, y_2 < y\in {\cal Y})\;
\left[\,\abs{c(y)}{\bar \varepsilon } + {u_0}\left(d(y)-c(y)A^{-1}b\right)\,<0\,\right]
\tag*{\mbox{(\bf H$^*$)}}
\\
&
(\exists \, y_1\in {\cal Y}) \;
(\forall \, y_1 > y\in {\cal Y}) \;
\left[\,-\abs{c(y)}{\bar \varepsilon } + {u_0} \left(d(y)-c(y)A^{-1}b\right)\,>0\,\right]
\notag
\end{align*}
(where $\abs{c}$ denotes the norm of the vector $c$)
and it says that the inequality in {\bf (H)} is preserved under small enough
perturbations proportional to $c(y)$, as long as $y$ is large or small
enough.

\bl{lemma:hstar-enough}
Suppose that {(\bf H)} and {(\bf H$^*$)} are satisfied.
Then, for each step input $u\equiv u_0\not= 0$, and every initial condition,
the corresponding solution $(x(t),y(t))$ of~(\ref{system}) is bounded.
\els

From Proposition~\ref{theorem-adapt} and Lemma~\ref{lemma:hstar-enough}, there
is the following immediate consequence:

\bc{cor:hstar-enough}
Suppose that {(\bf H)} and {(\bf H$^*$)} are satisfied.
Then, for each step input $u\equiv u_0\not= 0$, and every initial condition,
the corresponding solution $(x(t),y(t))$ of~(\ref{system})
converges to $(x_0,y_0)$.
\ecs

Condition {(\bf H$^*$)} is often automatically satisfied in examples:

\bl{lemma:affine}
Suppose that $c(y)$ and $d(y)$ are affine in $y$.  That is, there are 
two row vectors $c_0,c_1\in \R^{1\times n}$ and two scalars $d_0,d_1$ such that
$c(y)=c_0+ yc_1$ and $d(y)=d_0+yd_1$.  Then, Property {\bf (H)} implies
Property {\bf (H$^*$)}.
\els

\section{Approximate adaptation}
\label{sec:approximate}

Perfect adaptation is an ideal mathematical property.
In biological systems, regulated behavior may break down due to dilution,
turn-over due to gene expression and protein degradation, and other
effects, especially over long time intervals.
From a modeling viewpoint, it is thus interesting to study mechanisms which
provide ``approximate'' adaptation, in the sense that the response of the
system remains approximately constant, as long as parameters (kinetic
constants, production  rates, degradation rates) stay within appropriate
ranges.
The reader is referred to the papers~\cite{iglesias_eds1,iglesias_eds2}
for formulations of certain approximate adaptation mechanisms for linear and
nonlinear models.
This section discusses a few general facts, and works out details for a
particular class of systems, to be illustrated with an example in
Section~\ref{sec:RNA}.

One general fact is that a perturbation of the right-hand side of
a differential equation results in small perturbations of trajectories,
on bounded intervals of time.
Specifically, suppose that $x(t)$ is the solution of a set of differential
equations 
$\dot x=f(x)$, with initial condition $x(0)=x_0$, and consider any fixed
time interval $[0,T]$.  Next, consider a perturbed equation $\dot z=f(z)+h(z)$,
and let $z(t)$ be its solution with the same initial condition $z(0)=x_0$.
Then, if the vector field ``$h$'' is small in an appropriate sense (uniformly,
for instance, or more generally if its integral along trajectories is small),
then it follows that 
$z(t)\approx x(t)$ for all $t\in [0,T]$; 
see Theorem 55 in Appendix C of~\cite{mct} for details.
In principle, and in the absence of additional stability assumptions, the
theoretical estimates tend to be conservative, in that the guaranteed
approximation is very poor as 
$T$ increases.  However, in practice the approximation may be quite good.
As an illustration, consider once again the ``sniffer'' reactions 
(\ref{sniffer_equations}) from~\cite{sniffers}, and suppose that one perturbs
the right-hand side of the equations by adding saturated terms:
\beqn
\dot x &=& - k_4x + k_3u + \frac{V_1y}{K_1+y}\\
\dot y &=&  - k_2xy + k_1u + \frac{V_2x}{K_2+x}
\eeqn
representing cross-activating feedbacks.
Using the step input from Figure~\protect{\ref{fig:4inputs}}(a),
Figure~\ref{fig:ffpert} compares the results of simulations (starting from
the zero initial state) of the original and the perturbed systems,
when all constants have been chosen as 1.  Notice that the perturbed system
has a response which is quite close to that of the original system, on the
given interval.
\fighere{\begin{figure}[h,t]
\begin{center}
\includegraphics[scale=0.5]{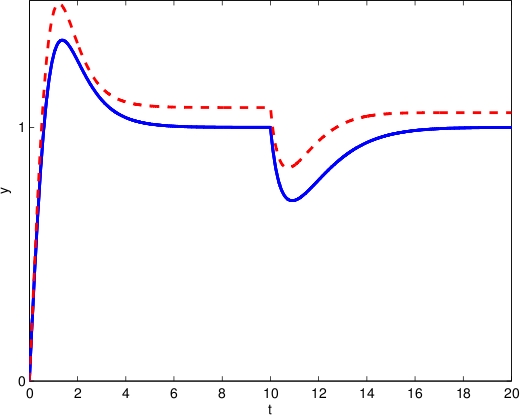}
\label{fig:ffpert}
\caption{Approximate adaptation: comparison of responses when right-hand side
  of equation is perturbed as discussed in text.  Input is as in
  Figure~\protect{\ref{fig:4inputs}}(a).  Solid line is for original system
  and dashed line for perturbed system.}
\end{center}
\end{figure}}

Another sense of approximate adaptation to step inputs is when adaptation
behavior happens only for input signal values in a restricted range.
This is illustrated next, using a perturbation of the ``sniffer'' reactions
(\ref{sniffer_equations}) from~\cite{sniffers}.
Suppose that the equations are as follows:
\begin{subequations}
\label{eq:mRNA}
\begin{eqnarray}
\dot x &=& - k_4x + k_3u + r(y)\\
\dot y &=&  - k_2xy + k_1u - p(y).
\end{eqnarray}
\end{subequations}
The term $p(y)$ may represent, for example, a linear or nonlinear degradation
effect for the species $y$, while $r(y)$ might represent an activating feedback.
(In Section~\ref{sec:RNA}, it will be shown that a microRNA-based feedforward
loop studied in the literature can be represented, after a coordinate change,
in this form.)
The steady state corresponding to a given constant input $u\equiv u_0$ can be found
as follows.  Setting the right-hand sides of~(\ref{eq:mRNA}) to zero and
solving the first equation for $x$ gives $x=\frac{1}{k_4}(k_3u + r(y))$.
The expression for $x$ is then substituted in the second equation, to provide
the following relation for $u$ in terms of $y$:
\[
u \;=\; Q(y) \;=\; \frac{k_2yr(y)+k_4p(y)}{k_1k_4-k_2k_3y}
\]
Suppose that $p(0)=0$ and that $k_2yr(y)+k_4p(y)$ is an increasing function of
$y$ (this happens automatically if, for example, 
$p(y)$ and $q(y)$ are non-negative and increasing functions of $y$).
Then $Q$ is an increasing function on the interval
$0\leq y<\alpha =\frac{k_1k_4}{k_2k_3}$, with $Q(0)=0$ and a pole at $\alpha $,
see Figure~\ref{fig:Q}.
\fighere{\begin{figure}[h,t]
\begin{center}
\includegraphics[scale=0.5]{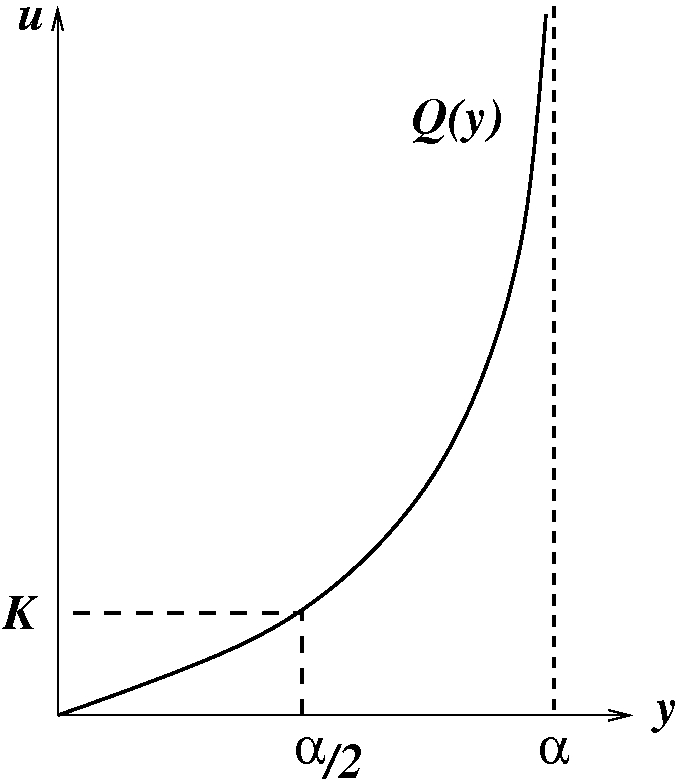}
\label{fig:Q}
\caption{The function $Q$}
\end{center}
\end{figure}}
The function $Q$ can be then inverted, so as to obtain $y$ as a function of $u$,
$y = Q^{-1}(u)$.  

For every step input $u$ whose amplitude is large enough, the steady state value
of the response $y$ is close to $\alpha $.
In that sense, ``approximate'' adaptation holds.
One way to characterize this effect is as follows.
Define $K = Q(\alpha /2)$.  This number plays a role analogous to that
of a ``half maximal effective
concentration'' or ``EC$_{50}$ value'' in pharmacology and biochemistry: for any
input value $u$ larger than $K$, $y$ is within $\alpha /2$ and $\alpha $.

For example, if $p(y)=ky$ (linear degradation/dilution) and $q(y)=0$, then
$y = Q^{-1}(u) = \frac{Vu}{K+u}$ for appropriate numbers $V$ and $K$.
For $u>K$, the response value is within $V/2$ and $V$.

\section{Pulse memory effects}
\label{sec:memory}

As remarked in the Introduction, when a pulse input $u(t)$ is applied to the
system~(\ref{system}), the asymptotic value of $y$ does not typically return
to its adapted value.  There is a ``memory'' effect as the asymptotic value of
$y$ depends on the magnitude of the step.
Thus, this section analyzes the effect of a pulse input, that is
$u(t)=u_0\not= 0$ for some interval $t\in [0,T]$, and $u(t)=0$ for $t>T$.
The underlying assumption is that the interval is long enough ($T\gg1$) so
that $x(T)$ and $y(T)$ are (approximately) in steady state:
$x(T)\approx-A^{-1}bu_0$ and $y(T)\approx y_0$.
This means that
$x(t) \approx -e^{(t-T)A}A^{-1}bu_0$ for $t\geq T$, and $y$ approximately solves
$\dot y = -c(y)e^{(t-T)A}A^{-1}bu_0$ starting at the adapted value $y_0$.

Therefore, and changing for simplicity the origin of time to $t=T$, one wishes
to estimate the limiting value of the solution of the initial-value problem:
\be{ivp}
\dot y \;=\; -c(y)e^{tA}A^{-1}bu_0\,,\quad
y(0)=y_0\,.
\ee
In general, such a differential equation, even though scalar, is not easy to
solve, because of the time dependence.
Two special cases are as follows.

\subsection{Affine case}

When $c(y)=c_0+yc_1$ is affine in $y$, one may write~(\ref{ivp}) as:
\[
\dot y + \alpha (t)y \;=\; \beta (t)
\]
where $\alpha (t)=c_1e^{tA}A^{-1}bu_0$ and $\beta (t) = -c_0e^{tA}A^{-1}bu_0$.
This is a linear differential equation, which can be solved in a standard
manner by using the integrating factor $e^{\int \alpha (t)dt}$.

\subsection{Separable case}

Another special case is that in which one may decompose $c(y)$ as follows:
\be{assumption:memory:case}
c(y) \;=\; \theta (y)c
\ee
where $\theta (y)$ is a nowhere vanishing scalar continuous function of $y$, and
$c$ is a row vector. 
Details in this special case are as follows.

Separating variables,
\[
\int_{y_0}^{y(t)} \frac{dz}{\theta (z)} \;=\;
-c \left( \int_0^t e^{sA}\,ds\right) A^{-1}bu_0
\;=\;
c\left(I-e^{tA}\right) A^{-2}bu_0 \,.
\]
The function $\Q_0(y):=\int_{y_0}^{y} \frac{dz}{\theta (z)}$ is strictly increasing
or strictly decreasing if $\theta $ is positive or negative respectively, so its
inverse $\Q:=\Q_0^{-1}$ is well defined, as it is also strictly increasing
or strictly decreasing respectively.

One concludes that:
\[
\hat y_{u_0} \;=\; \Q\left(cA^{-2}bu_0\right)
\]
is the steady-state value of $y$ after the system has been subjected to a long
pulse of amplitude $u_0$ and the pulse is removed.
Recalling the conditions under which $\Q$ is increasing or decreasing, one may
summarize as follows:

\bp{prop:memory}
Assuming the form in~(\ref{assumption:memory:case}), with $\theta (y)$ nonzero and
continuous on ${\cal Y}$, the steady state value $\hat y_{u_0}$ is:
\bi
\item[(a)]
an increasing function of $u_0$ if $\theta (y_0)cA^{-2}b>0$, and 
\item[(b)]
a decreasing function of $u_0$ if $\theta (y_0)cA^{-2}b<0$.
\ei
\eps

Notice that $\Q(0)=y_0$, so the steady state $\hat y_{u_0}$ after a pulse is
smaller (respectively, larger) than the adaptation steady state $y_0$ (that
results after a step input) provided that $\theta (y_0)cA^{-2}b<0$ (respectively,
$>0$).

\subsection{Non-ideal pulses}

Observe that, while the previous analysis concerned only ideal pulses
(the value of $u(t)$ returns exactly to zero after time $t=T$), approximately
the same phenomenon will still occur if $u$ merely returns to a ``small''
value, in the following sense.
Suppose that $u(t)=u_0\not= 0$ for some interval $t\in [0,T]$, and $u(t)=\varepsilon $ for
$t>T$, where $\varepsilon $ is sufficiently small.
The asymptotic value of the response $y$ will eventually converge to the
adapted value $y_0$, if $\varepsilon \not= 0$. 
However, if $\varepsilon \approx0$, this convergence to $y_0$ will be extremely slow, as
the behavior will be close of that for ideal pulses.
Mathematically, this fact is a trivial consequence of the continuous
dependence of solutions of differential equations on parameters (on finite
time intervals). 

To formulate this property precisely,
call $z(t)$ the value of $y(t)$ that would correspond to the ideal
pulse ($u(t)=0$ for $t>T$), and $y_\varepsilon (t)$ the value that corresponds
to the input with $u(t)=\varepsilon $ for $t>T$.   The subscript $\varepsilon $ is used to
emphasize the dependence on the numerical value of $\varepsilon $.
For any fixed time $T'>T$, it holds that $y_\varepsilon (T')\approx z(T')$,
in the sense that $y_\varepsilon (T')\rightarrow z(T')$ as $\varepsilon \rightarrow 0$.  (See for example,
Theorem 1 in~\cite{mct} for a proof; explicit estimates of convergence
can be obtained using the Gronwall inequality, as discussed in that textbook.)

Another way in which a pulse may differ from an ideal pulse is if the cut-off
to $u(t)=0$ is not sharp.
In fact, current microfluidics technologies allow one to produce pulsatile-like
inputs to cell signaling systems, but these might exhibit a slow decay at the
tail end.  (The same effect may manifest itself in real systems, when an
intermediate species stands between the input and the $x$ and $y$ variables.)
If the cut-off is sharp enough, a continuity argument similar to the one
explained for pulses that sharply return to a constant value $\not= 0$ applies.
However, for slower decays to $u=0$, the asymptotic value of the
response may indeed return to close to adapted values, as was earlier
illustrated with an example, see Figure~\ref{fig:4inputs}(h).

\section{Proofs}
\label{sec:proofs} 

The following is an elementary observation about scalar differential equations.

\bl{lemma:robustness}
Consider the scalar time-dependent differential equation
\[
\dot y \;=\; F(t,y) + \varphi(y)
\]
where $F$ and $\varphi$ are differentiable functions.
Assume that $F(t,y)\rightarrow 0$ as $t\rightarrow \infty $ uniformly on $y\in K$,
where $K\subseteq \R$ is a closed and bounded interval, and
that there is some $y_0\in K$ such that $\varphi(y)>0$ for all
$y<y_0$ and $\varphi(y)<0$ for all $y>y_0$.  Then, every solution
$y:[0,\infty )\rightarrow K$ is so that $y(t)\rightarrow y_0$ as $t\rightarrow \infty $.
\els

\bpr
Let $y(t)$ be a solution with values in $K$, and pick any open neighborhood $N$
of $y_0$.  We must show that, for some $T$, $y(t)\in N$ for all $t>T$.
The set $K\setminus N$ is the union of two closed and bounded sets $A_-$
and $A_+$ (either of which might possibly be empty, if $y_0$ is an endpoint
of $K$) such that $y\in A_-\Rightarrow y<y_0$ and $y\in A_+\Rightarrow y>y_0$.  By continuity of
the function $\varphi$, there is some positive number $\delta $ such that
$\varphi(y)>\delta $ for all $y\in A_-$ and $\varphi(y)< -\delta $ for all $y\in A_+$.
For some $t_0$, $\abs{F(t,y)}<\delta /2$ for all $y\in K\setminus N$ and all $t\geq t_0$.
Thus, for $t\geq t_0$ we have that $\dot y(t)>\delta /2$ if $y(t)\in A_-$ and
$\dot y(t)< -\delta /2$ if $y(t)\in A_+$.  This means that for some $T>t_0$ it will
hold that $y(t)$ exits $K\setminus N$ and does not enter again, as needed.
\epr

\subsubsection*{Proof of Proposition~\protect{\ref{theorem-adapt}}}

We will apply Lemma~\ref{lemma:robustness}.
Suppose that Property {\bf (H)} holds.
Pick any step input $u\equiv u_0\not= 0$ and an initial condition
of~(\ref{system}), and suppose that the corresponding solution
$(x(t),y(t))$ of~(\ref{system}) is so that $y(t)$ is bounded.
Since ${\cal Y}$ is a closed set, this is the same as saying that
$y(t)\in K$ for all $t$, for some closed and bounded interval $K\subseteq {\cal Y}$.
We have that
$x(t)=\theta (t)-A^{-1}bu_0$ for all $t\geq 0$, 
where $\theta (t)=e^{tA}\left(x_0 + A^{-1}bu_0\right)$
and
therefore $y$ satisfies:
\be{eq:theorem:eq1}
\dot y \;=\; F(t,y) + \varphi(y) \;=\; 
c(y)
\theta (t)
+ \varphi(y)
\ee
where $\varphi(y)=\left(d(y)-c(y)A^{-1}b)\right) u_0$.
Property~{\bf (H)} gives the property needed for $\varphi$ in the Lemma.
On the other hand, $\theta (t)\rightarrow 0$ (as $e^{tA}\rightarrow 0$, by stability), so
$F(t,y)\rightarrow 0$ uniformly on $y\in K$.
Thus, Lemma~\ref{lemma:robustness} gives the desired conclusion.
\qed

\subsubsection*{Proof of Lemma~\ref{lemma:hstar-enough}}

We pick any initial condition, and want to show that the corresponding
solution $(x(t),y(t))$ of~(\ref{system}) 
is such that $y(t)$ is bounded.
We will prove that $y(t)$ is upper-bounded, the lower bound proof being
similar, and
assume that ${\cal Y}$ is not upper bounded, since otherwise we are done.
By Property {\bf (H$^*$)}, we may pick ${\bar \varepsilon >0}$ and $y_2\in {\cal Y}$ such that
$
\abs{c(y)}{\bar \varepsilon } + \varphi(y)<0
$
for all $y>y_2$, where the function $\varphi$ is as in the proof of
Proposition~\ref{theorem-adapt}.
By~(\ref{eq:theorem:eq1}), we know that $\dot y(t)<0$ as long as $y(t)>y_2$
and $t\geq t_0$, where $t_0$ is picked so that $\abs{\theta (t)}<\bar{\varepsilon }$ for all
$t\geq t_0$.
This clearly implies that $y(t)$ is upper bounded.
\qed

\subsubsection*{Proof of Lemma~\ref{lemma:affine}}

Suppose that $c(y)$ and $d(y)$ are affine in $y$,
$c(y)=c_0+ yc_1$ and $d(y)=d_0+yd_1$.
As Property~{\bf (H)} holds,
\[
\mu  + \nu y \,=\,d_0+yd_1 - (c_0+ yc_1)A^{-1}b \,<\,0
\]
for $y>y_0$, $y\in {\cal Y}$,
where we are writing
$\mu :=d_0-c_0A^{-1}b$ and $\nu :=d_1-c_1A^{-1}b$.

Fix any $u_0>0$.
We need to show that Property {\bf (H$^*$)} holds.
We will show the existence of $\bar\varepsilon $ and $y_2$;
existence of $y_2$ is proved in a similar way.
Our goal is to pick 
$y_2\in {\cal Y}$
in such a way that
\be{ineq:H*:proof:0}
\abs{c(y)}\bar{\varepsilon } + {u_0}\left( d(y) - c(y)A^{-1}b\right) \,<0
\ee
whenever $y> y_2$ is an element of ${\cal Y}$.
If the interval ${\cal Y}$ is upper bounded, we may pick 
$y_2$ equal to its right endpoint,
and this property is satisfied vacuously.
Thus, we assume from now on that ${\cal Y}$ is not upper bounded.

We claim that $\nu <0$.  Indeed, if $\nu \geq 0$ then $\mu  + \nu y\geq \mu +\nu y_0=0$ for 
any $y>y_0$, $y\in {\cal Y}$,
which would contradict Property~{\bf (H)}. (Note that there exist such
$y>y_0$, because $y_0$ cannot be the right endpoint of ${\cal Y}$, because ${\cal Y}$ is
not upper bounded.)

For~(\ref{ineq:H*:proof:0}) to be satisfied, and assuming we pick $y_2\geq 0$,
it is enough that this inequality should hold:
\be{ineq:H*:proof}
\left(\abs{c_0}{\bar\varepsilon }+u_0\mu \right)+y\left(\abs{c_1}{\bar\varepsilon }+u_0 \nu \right)\,<0
\ee
(because $\abs{c_0+c_1y}\leq \abs{c_0}+\abs{c_1}y$).
We let $\bar{\varepsilon }:=-\frac{\nu }{2\abs{c_1}u_0}$.
Then, for $y>0$,
\[
\left(\abs{c_0}{\bar\varepsilon }+u_0\mu \right)+y\left(\abs{c_1}{\bar\varepsilon }+u_0 \nu \right)
\;<\;
\left(\abs{c_0}{\bar\varepsilon }+u_0\mu \right)+yu_0 \nu /2
\]
and, since the upper bound is a linear function with negative slope,
it will be negative for large $y$.
\qed

\section{Examples}
\label{sec:examples}

We show here how the results apply, in particular, to the models in
the papers~\cite{sniffers} and ~\cite{iglesias05}.

\subsection{``Sniffer'' model}

The ``perfect adaptation'' model in~\cite{sniffers} is, after a renaming of
variables and a slight rearranging to bring into the form~(\ref{system}),
as shown in~(\ref{sniffer_equations}).
Here $y_0=\frac{k_1k_4}{k_2k_3}$, and {\bf (H)} is satisfied because
$d(y)-c(y)A^{-1}b = k_1-\frac{k_2k_3}{k_4}y$ changes sign at $y=y_0$.
This example has the form in Lemma~\ref{lemma:affine}, so convergence holds.

To study the effect of pulses, we write $c(y)$ in the 
form~(\ref{assumption:memory:case}) using $c=-k_2$ and $\theta (y)=y$.
So $\Q_0(y)=\int_{y_0}^y dz/z = \ln(y) - \ln(y_0) = \ln(y/y_0)$.
It follows that $\Q(r)=y_0e^r$, so
\[
\hat y_{u_0} \;=\;
y_0 \exp\left(-\frac{k_2k_3}{k_4^2} u_0\right)
\]
which decreases with $u_0$.

\subsection{\emph{Dictyostelium} chemotaxis models from~\protect{\cite{iglesias05}}}

There are several models given in~\cite{iglesias05}, but they all have the
same general interpretation.  The authors
of~\cite{iglesias05}, based on previous work~\cite{levchenko_iglesias02},
postulate the existence of a ``response regulator'' R, a variable that
correlates to the chemotactic activity of the system, that can be in an
``active'' or in an ``inactive'' form.  The activation and inactivation of R
are regulated by a pair of opposing processes: an excitation process that
induces an increase in the level of the response R, and an inhibition process
that lowers this response.
The input to the system is the extracellular chemoattractant concentration,
and it is assumed that this signal triggers increases in concentrations in
both the activation and inactivation elements.  We denote by $y(t)$ the
concentration of active regulator, by $\alpha -y(t)$ the concentration of inactive
regulator, where $\alpha $ is the total concentration (active$+$inactive), assumed
constant, and by $x_1(t)$ and $x_2(t)$ the concentrations of the activation
and inactivation elements respectively.

There are several alternative models given in~\cite{iglesias05}.
The first one is, in our notations:
\beqn
\dot x_1&=& -k_1 x_1 + k_2 u\\
\dot x_2&=& -k_3 x_2 + k_4 u\\
\dot y  &=& k_5(\alpha -y)x_1-k_6 y x_2
\eeqn
where the $k_i$'s are positive constants.
This has the form~(\ref{system}), with ${\cal X}=\R^2_{\geq 0}$, ${\cal Y}=[0,\alpha ]$,
$c(y)=(k_5(\alpha -y),-k_6y)$, and $d(y)=0$.
To simplify notations, let us write $\A=\frac{k_2k_5}{k_1}$ and
$\B=\frac{k_4k_6}{k_3}$.
Note that Property~{\bf (H)} is satisfied, as the solution $y_0$ of
\be{equation:iglesias-example1-ss}
\A(\alpha -y)-\B y = 0
\ee
belongs to the interval $(0,\alpha )$, and the algebraic expression changes there
from positive to negative.  The
$y$-dynamics is bounded, by definition, so we have convergence. 

Since the $y$ equation is affine, the memory effect of pulses can be obtained
by solving an appropriate linear differential equation, as explained earlier,
but the expression for the solution is algebraically very involved.
We can, however, make some qualitative remarks.

\bl{lemma:iglesias-example1}
The solution of the initial-value problem
\be{equation:iglesias-example1-time}
\dot y\;=\;\A(\alpha -y)e^{-k_1t}u_0 - \B ye^{-k_3t}u_0\,,\;\;
y(0)=y_0
\ee
has a limit $\hat y_{u_0}$.
If $k_3>k_1$ then $\hat y_{u_0}>y_0$, 
and if $k_3<k_1$ then $\hat y_{u_0}<y_0$.
\els

\bpr
We assume that $k_3>k_1$; the case $k_3<k_1$ is proved in an analogous fashion.
Since $y(t)$ is bounded, it will be enough to show that there cannot exist
two limit points $0\leq \bar y_1<\bar y_2\leq \delta $ of the solution.  So assume that
such points exist, and let $0<t_1<s_1<t_2<s_2\ldots  \rightarrow \infty $ be so that
$y(t_i)\rightarrow \bar y_1$ as $i\rightarrow \infty $ and
$y(s_i)\rightarrow \bar y_2$ as $i\rightarrow \infty $.
Pick $\varepsilon :=\frac{1}{2}(\bar y_2-\bar y_1)+(\alpha -\bar y_2)>0$ and some
time $\bar t>0$ such that
$\A\varepsilon >\alpha \B e^{(k_1-k_3) \bar t}$ (there is such a $\bar t$ because $k_1-k_3<0$).

Note the following property: 
\be{eq:iglesias-example1:eq1}
y(t)<\alpha -\varepsilon  \;\mbox{and}\; t\geq \bar t \;\Rightarrow \; \dot y(t)>0 \,.
\ee
Indeed, 
\[
\frac{e^{k_1t}}{u_0}\dot y \;=\;
\A(\alpha -y(t)) - \B y(t)e^{(k_1-k_3) \bar t} \;> \;
\A \varepsilon  - \B \alpha  e^{(k_1-k_3) t} \;> \; 0
\]
because $\alpha -y(t)>\varepsilon $ and $y(t)<\alpha $, so $\dot y(t)>0$ as claimed.

Since $\bar y_2>\alpha -\varepsilon $, there is some $i$ so that $y(s_i)>\alpha -\varepsilon $.
Without loss of generality, we may assume that $s_i$ was picked larger than
$\bar t$.
Then,
\be{eq:iglesias-example1:eq2}
t>s_i \;\Rightarrow \; y(t)\,\geq \,\alpha -\varepsilon  \,.
\ee
(Otherwise, there would exist some $a<\alpha -\delta $ and some $T>s_i$ such that
$y(T)=a$, and we can assume that $T$ has been picked smallest possible with
this property, for the given $a$.
Pick $\delta >0$ so that $s_i<T-\delta $ and so that the interval $I=(T-\delta ,T]$
has the property that $y(I)\subseteq [0,\alpha -\varepsilon )$.  
Then, by property~(\ref{eq:iglesias-example1:eq1}), $\dot y(t)>0$ for all $t\in I$.
So $y(t)<y(T)$ for all $t\in I$, which contradicts the minimality of $T$.)

On the other hand, since $\bar y_1<\alpha -\varepsilon $, $y(t_j)<\alpha -\varepsilon $ for all sufficiently
large $j$.  This contradicts~(\ref{eq:iglesias-example1:eq2}) if $j>i$.
The contradiction shows that such $\bar y_1<\bar y_2$ cannot exist, so the
function $y(t)$ is convergent as $t\rightarrow \infty $.

We must now prove that $\hat y_{u_0}>y_0$.
Since $y_0$ solves~(\ref{equation:iglesias-example1-ss}), and
$e^{(k_1-k_3)t}<1$ for all $t\geq 0$, it follows that
$\A(\alpha -y_0)e^{-k_1t}-\B y_0e^{-k_3t}>0$ for all $t\geq 0$.
Thus:
\[
\dot y(t) 
\;=\;
\A(\alpha -y(t))e^{-k_1t} - \B y(t)e^{-k_3t} 
\;\geq \;
\A(\alpha -y_0)e^{-k_1t} - \B y_0e^{-k_3t} 
\;>\;
0
\]
whenever $t\geq 0$ is such that $y(t) \leq  y_0$.  This implies that $y(t)>y(0)=y_0$
for all $t>0$.  Therefore the limit also satisfies $\hat y_{u_0}>y_0$.
\epr

The interpretation of Lemma~\ref{lemma:iglesias-example1} is obvious:
$k_3>k_1$ means that the inhibitor ($x_2$) degrades at a faster rate than
the activator ($x_1$).  Thus, when the external signal is turned-off, there is
a residual effect due to the additional activator still present, which implies
a positive memory effect, in the sense that the response is higher than its
value at $u=0$.  Similarly, when $k_3<k_1$, there is additional
repressor present and the memory effect is negative (that is, lower than
when $u=0$).

The other models from~\cite{iglesias05} are similar, differing only in the
placement of the feedforward terms in the $y$ equation, and the stability
results apply equally well.
Let us consider one of the variants:
\beqn
\dot x_1&=& -k_1 x_1 + k_2 u\\
\dot x_2&=& -k_3 x_2 + k_4 x_1\\
\dot y  &=& k_5(\alpha -y)u-k_6 y x_2 \,.
\eeqn
Note that the activator now acts so as to enhance the inhibitor, and the
input signal acts directly on the response element.
This has the form~(\ref{system}), with ${\cal X}=\R^2_{\geq 0}$, ${\cal Y}=[0,\alpha ]$,
$c(y)=(0,-k_6y)$, and $d(y)=k_5(\alpha -y)$.
Note that Property~{\bf (H)} is satisfied, as 
${k_5}(\alpha -y)-\frac{k_2k_4k_6}{k_1k_3}y$ changes sign at a
$y_0\in (0,\alpha )$.  The $y$-dynamics is bounded, by definition, so we have
convergence. 

Once again, since the $y$ equation is affine, the memory effect of pulses can
be computed using linear differential equations theory.
\emph{The memory effect is in this case decreasing in $u_0$, independently of
  the parameters $k_i>0$.}
This is because the initial value problem after the signal has been turned-off
has the form:
\[
\dot y(t) \;=\; - y(t) \theta (t)
\]
for some positive function $\theta (t)$, so $y(t)<0$ for all $t$.
Intuitively: the activation effect ($u$) turns-off immediately, but there
is a residual inhibition effect ($x_2$).

As an concrete illustration, let us work out the case in which all constants
are equal to $1$.  In this case, solving $(1-y)-y=0$
gives $y_0=1/2$, and $\tilde y_{u_0}$ is the limiting value of the solution of:
\[
\dot y \;=\; -ye^{-t}(t+1)u_0
\]
(because $x_2(t)=e^{-t}(t+1)u_0$).
Thus, $y(t)=\frac{1}{2}e^{[(2+t)e^{-t}-2]u_0}$
and therefore $\hat y_{u_0} = \frac{1}{2}e^{-2u_0}$,
so that, indeed, the memory effect is negative.

\subsection{A Michaelis-Menten model}

Non-affine variant of the above examples may be obtained by using
Michaelis-Menten dynamics for activation and inhibition reactions.
For instance, we may write:
\[
\dot y \;= \; \frac{V_1(\alpha -y)}{K_1+(\alpha -y)} u -\frac{V_2y}{K_2+y} x_2 
\]
for some positive constants $V_i$ and $K_i$.
Once again, our hypotheses apply, and there is convergence to $y_0$.

We study memory effects for pulses for this example, but only in the special
case in which all constants are equal to 1 and for a 1-dimensional $x$-system:
\beqn
\dot x &=& -x+u\\
\dot y &=&  \frac{(1-y)}{1+(1-y)} u - \frac{y}{1+y} x \,.
\eeqn
We have that $y_0=1/2$ and $x_0=u_0$, and $c(y)=-y/(1+y)$.
We are in the separable situation described earlier, and so solve
$\dot y = c(y)e^{-t}u_0$ using separation of variables.
Writing 
\[
\ln y(t) + \ln 2 + y(t) - 1/2\;=\;
\int_{1/2}^{y(t)} (\frac{1}{s} +1)\,ds \;=\; -\int_0^te^{-s}\,ds\,u_0
\;=\; (e^{-t}-1)u_0
\]
we have, taking limits, that $\hat y_{u_0}$ is the solution of the
algebraic equation:
\[
\ln \hat y_{u_0} + \hat y_{u_0} = 1/2 - \ln 2 - u_0
\]
and so decreases with $u_0$.
For example, if $u_0=1$ then $\hat y_{u_0}\approx0.24$, which is less than
one-half of the adapted value $y_0=0.5$.

\subsection{A carbohydrate uptake system}

The paper~\cite{Kremling} deals with a feedforward motif that appears in one of
the carbohydrate uptake systems in \emph{E.coli}, the phosphotransferase
system (PTS) for glucose uptake involving phosphoenolpyruvate (PEP).
The reactions in this system result in the phosphorylation of Enzyme II A.
``EIIA-P'' is used to denote the phosphorylated form of this enzyme,
which in turn has various regulatory functions through synthesis of cAMP.
A feedforward loop is obtained when viewing $u=$ Glc6P
(glucose 6-phosphate) as input to the system, and using as state variables
the concentrations $x=$ TP (triose phosphate) and $y=$ PEP.  See Figure 2 in
that paper: there are positive effects of $u$ on $x$, and of $x$ on $y$, as
well as a countering negative direct effect of $u$ on $y$ that involves the
dephosphorylation of PEP into Prv via pyruvate kinase.
For a simplified analysis following~\cite{Kremling}, assuming equation
(1) from that paper, there results that the output of the system, the
concentration of EIIA-P, is an increasing function of the ratio of
concentrations PEP/Prv, where Prv is pyruvate (equations (3) and (5) in the
citation). 
The objective of keeping EIIA-P approximately constant is achieved if the
ratio PEP/Prv is kept approximately constant.
The model in~\cite{Kremling}, Supplemental Materials, equations (2,3,11)
together with the assumption that equation (1) in the paper holds, provides
the following equations for $x$ and $y$:
\begin{subequations}
\label{PEPmodel}
\begin{eqnarray}
\dot x &=& -ax + bu + \alpha \\
\dot y &=& ax - puy - \beta 
\end{eqnarray}
\end{subequations}
assuming that the simplest mass-action model is used for supplemental equation
(12) of the reference. 
The constants $\alpha $ and $\beta $ represent uptake rates (supplementary equations
(8,9)).
In the case when $\alpha =\beta =0$, there results a system of the general
form~(\ref{system}), with $A=a$, $c(y)=a$ and $d(y)=-py$ (affine case).
Solutions for constant nonzero inputs have $y(t)\rightarrow y_0 = b/p$, and the effect
of pulses can be analyzed easily.
When $\alpha $ and $\beta $ are nonzero, the adaptation property fails, although
it holds approximately if these numbers are small.
However, even for large $\alpha ,\beta $, the steady state when $u$ is a constant
equal to $u_0$ is:
\[
y_0 \;=\; \frac{b}{p} \,+\, \frac{\alpha -\beta }{pu_0}\,.
\]
This is still approximately constant, $y_0\approx b/p$, provided that
$u_0$ be sufficiently large, just as in the previous discussion of approximate
adaptation. 

\subsection{An example of approximate adaptation}
\label{sec:RNA}

The paper~\cite{xu_mRNAfeedforward09} provides the following model of
microRNA-mediated feedforward adaptation:
\beqn
\dot p_1 &=& \alpha _1\omega  - \beta _1p_1\\
\dot m_1 &=& \frac{\alpha _2p_1^m}{1+p_1^m} - \gamma  m_1m_2 - \beta _2m_1\\
\dot m_2 &=& \frac{\alpha _3p_1^m}{1+p_1^m} - \gamma  m_1m_2 - \beta _3m_2\\
\dot p_2 &=& \alpha _4m_2 - \beta _4p_2
\eeqn
where $p_1, m_1, m_2, p_2,\omega $ are respectively the species concentrations of an
``upstream factor'', a microRNA, a target mRNA, the protein produced by the
target mRNA, and an inducer of the upstream factor.  The various constants
represent transcription, translation, and degradation rates as well as well as
the efficiency of pairing of the microRNA to its target.  (As in the
reference, we pick identical Hill coefficients for both promoters.)
The interest in~\cite{xu_mRNAfeedforward09} is in studying the robustness of
the steady state value of $p_2$.  Since this value is directly proportional to
the steady state value of $m_2$, we omit $p_2$ from the model from now on.
Similarly, as $\frac{p_1^m}{1+p_1^m}$ is an increasing function of $p_1$,
which is in turn proportional to $\omega $ at steady state, we will think of this
term as an input ``$u$'' and drop the equation for $p_1$ as well.
We are left with the following two-dimensional $(m_1,m_2)$ system:
\beqn
\dot m_1 &=& \alpha _2u - \gamma  m_1m_2 - \beta _2m_1\\
\dot m_2 &=& \alpha _3u - \gamma  m_1m_2 - \beta _3m_2.
\eeqn
When expressed in the alternative coordinates $x=m_1-m_2$ and $y=m_2$,
the system has the form in Equations~(\ref{eq:mRNA}), with
$k_1=\alpha _3$,
$k_2=\gamma $,
$k_3=\alpha _2-\alpha _3$,
$k_4=\beta _2$,
$r(y)=(\beta _3-\beta _2)y$,
and $p(y)=\gamma y^2+\beta _3y$.
Thus this system exhibits an approximately adaptive behavior for large inputs,
as discussed in Section~\ref{sec:approximate}.
In particular, consider the parameter values used
in~\cite{xu_mRNAfeedforward09}:
$\alpha _1=0.01$, 
$\alpha _2=0.1$,
$\alpha _3 =0.02$, 
$\alpha _4=0.01$, 
$\beta _1=0.001$, 
$\beta _2=0.0025$, 
$\beta _3=0.002$, 
$\beta _4=0.001$, 
and $\gamma _=0.001$.
Then $Q(y)=\frac{0.000002y^2+0.000005y}{0.00005-0.00008y}$
and, with the notations in Section~\ref{sec:approximate}, the ``adaptation''
value for $m_2$ is $\alpha =0.625$
with the ``EC$_{50}$ value'' $K=Q(\alpha /2)\approx0.07$.
The corresponding steady state value of $p_2$ is
$\alpha _4m_2/\beta _4 =6.25$.
(Compare Figure 2 in~\cite{xu_mRNAfeedforward09}.)

\section{Remarks on integral feedback}
\label{sec:imp}

The ``internal model principle'' (see e.g.~\cite{imp03}) states that, if a
system perfectly adapts to all step inputs, then it may be re-written,
possibly after performing a nonlinear change of coordinates, as a system in
which an integral of the regulated quantity (response variable) is fed-back.
A feedforward (not feedback) system that exhibits
perfect adaptation might appear at first sight to be a counter-example to this
fact. 
However, there is not necessarily a contradiction, as a change of coordinates
may allow one to transform a feedforward into a feedback system.
This observation was made in~\cite{iglesias_ejc03}, and we discuss it further
here through an example.

As an illustration, consider the following two-dimensional linear system: 
\begin{subequations}
\label{ff-to-be-fb}
\begin{eqnarray}
\dot x &=& -x+u\\
 \dot y &=& -x-ay+u
\end{eqnarray}
\end{subequations} 
(where ``$a$'' is some positive constant).  This system
has the property, when the input $u$ is constantly
equal to a value $u_0$, that every solution converges to the state
$(u_0,0)$ as $t\rightarrow \infty $.  Thus, the response variable $y(t)$ converges to $y_0=0$
no matter what is the actual value of $u_0$.   The system response is
perfectly adaptative.

The system~(\ref{ff-to-be-fb}) has a feedforward form.
However, the same system can be recast as an integral feedback
system, as follows.
Suppose that we choose to represent the system using the state variables
$z=x-y$
and $y$ instead of $x$ and $y$.  In the new set of coordinates:
\begin{subequations}
\label{ff-now-fb}
\begin{eqnarray}
 \dot z &=& ay\\
 \dot y &=& -z - (a+1)y  + u
\end{eqnarray}
\end{subequations} 
which can be viewed as a system in which the rate of change of the regulated
quantity $y$ depends on $y$ itself (proportional negative feedback) as well as
on $z$, which is (up to a positive constant multiple) the integral of $y$
(integral feedback term).

Notice that, especially when seen as an integral feedback system, it is
immediately obvious that for \emph{every} step input $u\equiv u_0$ (not merely
nonzero steps), any steady state has the value $y=0$, since $0=\dot z=ay$ at
steady states.  So, after a pulse, the system will eventually also
converge to the adapted value (since we can see the behavior, after the end of
the pulse, as the behavior corresponding to a zero step).  Thus, the memory
effect discussed in this note will not occur for a true integral feedback
system such as the linear system shown above.
(Note that the system~(\ref{ff-to-be-fb}) does not have the exact
form~(\ref{system}) studied in this paper, because of the
 additive, not multiplicative, term ``$-ay$''.)

As an example, the plot shown in Figure~\ref{fig:fffb}(b) shows
the response to the pulse in Figure~\ref{fig:fffb}(a),
for system~(\ref{ff-to-be-fb}) with $a=0.1$.
Adaptation to $y=0$ results in this case, which also happens with the response
in Figure~\ref{fig:fffb}(d) to the step input in Figure~\ref{fig:fffb}(c).
(Scales for $y$ have been normalized, so as to show relative changes.
The response in (d) eventually settles back to zero, not shown.)
Note that the adaptation to the pulse is faster than that for the step input.
\fighere{\begin{figure}[h,t]
\begin{center}
\includegraphics[scale=0.7]{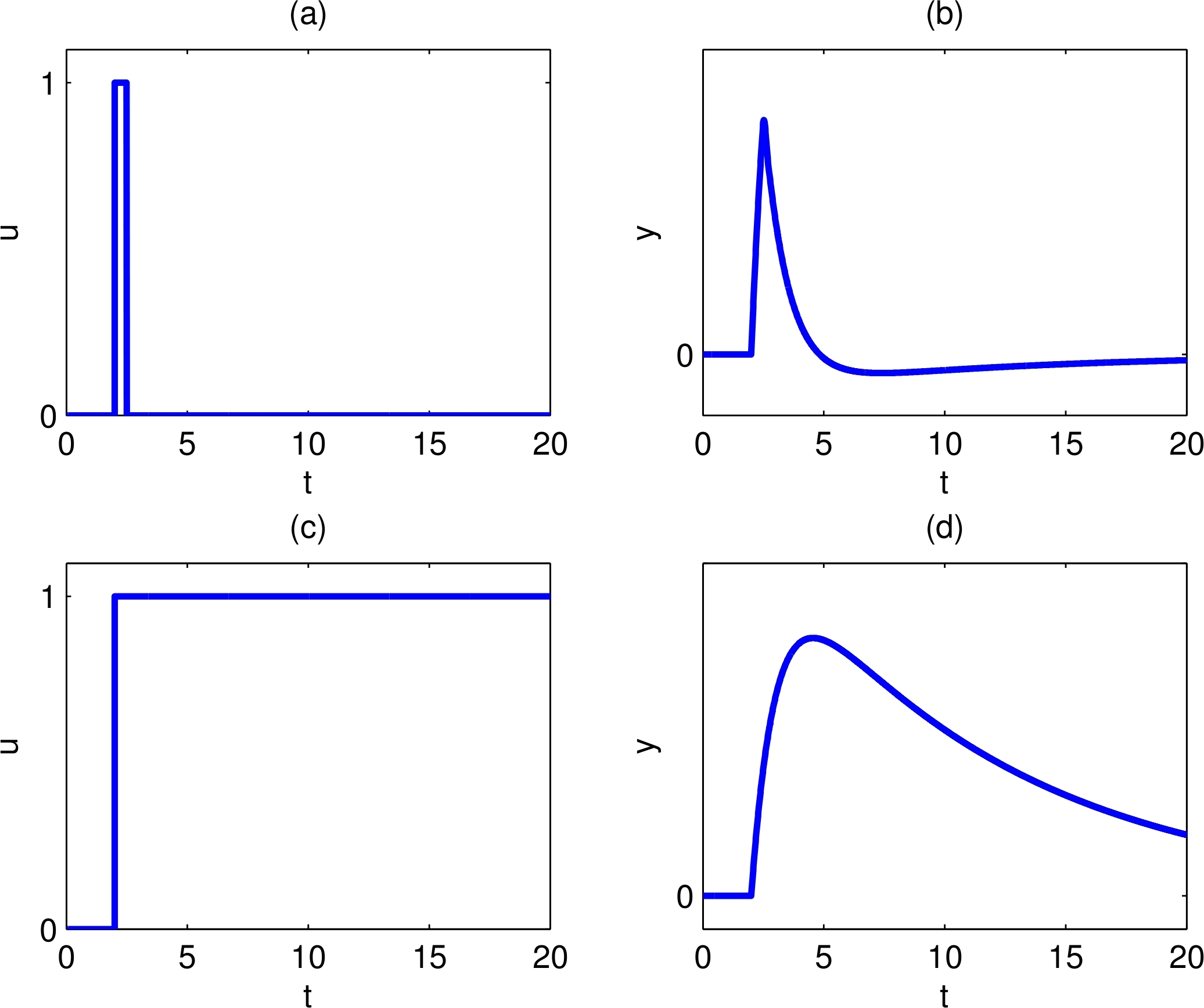}
\label{fig:fffb}
\caption{Responses to a pulse and to a step input, for a linear feedforward
  system discussed in the text.  (a) Pulse input.  (b) Response to the input
  in (a).  (c) Step input.  (d) Response to the input in (c).}
\end{center}
\end{figure}}
Interestingly, this model reproduces qualitatively Figure 2
from~\cite{bodenschatz}%
\footnote{The author thanks Pablo Iglesias for pointing out this reference.},
reproduced as Figure~\ref{fig:bodenschatz} in this paper.
The figure compares the changes in translocation of CRAC
(cytosolic regulator of adenylyl cyclase), reported by relative fluorescence
of a  CRAC-GFP construct, in chemotactic \emph{Dictyostelium} in response to
a ``short'' (i.e., pulse) or a ``continuous'' (i.e., step) stimulus generated
of cAMP.
\fighere{\begin{figure}[h,t]
\begin{center}
\includegraphics[scale=1]{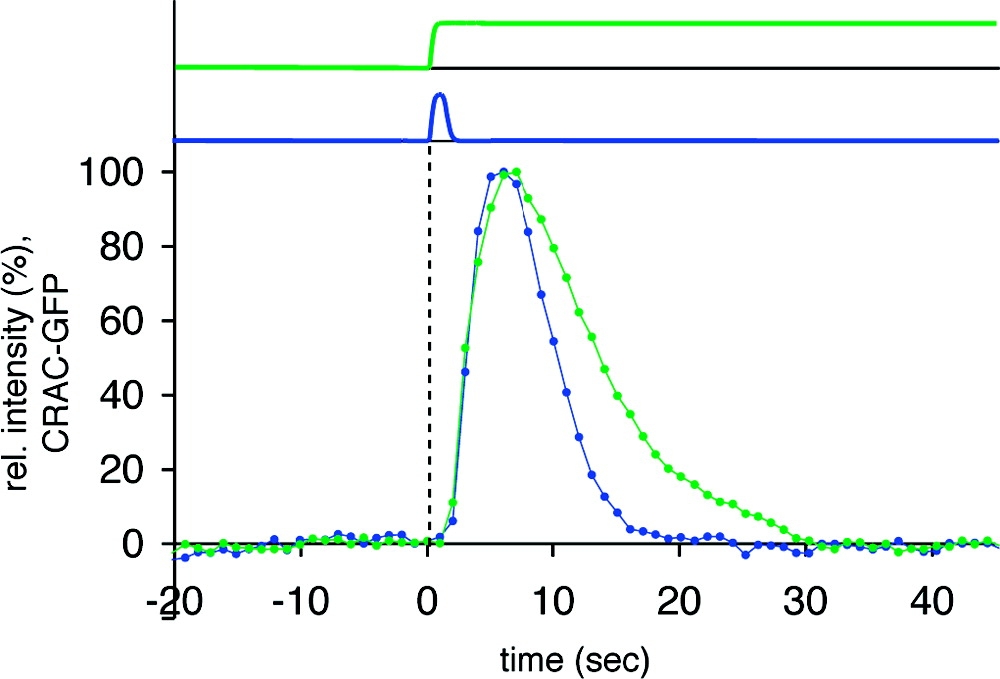}
\label{fig:bodenschatz}
\caption{Reproduced (microscopy inset removed) with permission from Figure 2
  in~\protect{\cite{bodenschatz}}.  Responses to step and pulse cAMP input
 in \emph{Dictyostelium}, as discussed in text.  The response to the pulse
 input settles faster than the response to the step.}
\end{center}
\end{figure}}

The integral feedback form~(\ref{ff-now-fb}) is often said to be more
``robust'' for adaptation than the feedforward form~(\ref{ff-to-be-fb}),
because the steady-state response $y$ is still zero even if the second
equation is arbitrary modified: if $\dot z=0$ and $a\not= 0$, one has that
$y=0$.  In contrast, modifications in~(\ref{ff-to-be-fb}) affect the steady
state value of $y$.  This claim of robustness is very misleading, however,
because perturbations of the \emph{first} equation in~(\ref{ff-now-fb}) will
generally change the steady state value of $y$.

The connection to the theory in~\cite{imp03} is somewhat subtle.  It is shown
there that, under appropriate technical restrictions on the dynamics, even a
nonlinear system that adapts to all step inputs can be recast as an integral
feedback system.  The key is the assumption ``all inputs'' -- the recasting
may fail to be global when inputs are restricted.  Rather than explaining here
the nonlinear theory, a simple local (linearized) version is analyzed next as an
illustration of these ideas.

Once again, take as an illustration the ``sniffer'' equations
from~\cite{sniffers} given in (\ref{sniffer_equations}).  For simplicity of
notations (nothing much changes in the general case), take all kinetic
constants equal to one:
\beqn
\dot x &=& - x + u \\
\dot y &=& - yx + u \,.
\eeqn
Suppose that one is only interested in studying the behavior of this system
in the vicinity of the steady state $(\bar x,\bar y)=(a,1)$ and the step input
$u(t)\equiv a$ for all $t$.
For small changes in initial states and input values, the system is
well-approximated by its linearization around these values, that is,
the system that is obtained when replacing
the nonlinear term ``$yx$'' by 
\[
(\partial (yx) / \partial x)x + (\partial (yx) / \partial y)y,
\]
where the partial derivatives are understood as evaluated at $(\bar x,\bar y)$.
Since
$\partial (yx) / \partial x =  {\bar y} = 1$ and
$\partial (yx) / \partial y =  {\bar x} = a$,
we have that the linearized system is precisely the system~(\ref{ff-to-be-fb}).
Thus, so long as $a\not= 0$, the system can be recast (locally) as an integral
feedback system.
However, in the special case when $a=0$, the recast system has the form
$\dot z = 0$, $\dot y = -z - y  + u$.
This system is \emph{not} an integral feedback system, since $z$ no longer
contains information about the integral of $y$.
(Mathematically, there is now a zero eigenvalue; thus, the system is no
longer asymptotically stable, and hence no regulation property holds.)

\section{Discussion}
\label{sec:discussion}

Adaptation is a feature often exhibited by biological systems, as discussed in
the cited references.  This paper started from the well-known observation that
certain types of feedforward circuits proposed in biological models have
adaptation properties, and established rigorous mathematical results along those
lines.

Perhaps of more interest, it was shown that a ``memory'' effect is often
displayed after pulses.  The magnitude of this effect is a nonlinear function
of the magnitude of the pulse, and estimates were given of its value.

One may speculate regarding what beneficial roles these memory effects of
pulses might play.  In at least some of the examples, the calculations show
that after a very high-amplitude pulse is turned-off, the response settles
down to a close to ``relaxed'' steady state.  In the context of a complex
system, this 
response might be appropriate, for example, in a situation in which resources,
used up while responding to a large external input, need to be replenished,
and this is achieved by turning-off processes controlled by the feedforward
circuit; in this way, a ``refractory period'' would be established.

Of course, the ``memory'' effect of pulses may or may not play a role in real
systems, because parameter ranges may be such that the effect is negligible,
or because sharp cutoffs of signals are rare in nature.  It remains to see if
feedforward circuits function in these regimes, in any real systems.

From an experimental viewpoint, the results in this paper suggest that one
might be able to use the pulse-memory property as a way to experimentally
distinguish true integral feedback systems from feedforward ones, through
the testing of system responses against ideal pulses.

\section*{Acknowledgments}

The author wishes to thank Yuan Wang and Pablo Iglesias, as well as two
anonymous reviewers, for many useful comments on the manuscript.
This research was supported in part by grants from AFOSR, NSF, and NIH,
and was carried out in part while visiting the Laboratory for Information and
Decision Systems at MIT.

\edo

\newpage

\begin{figure}[h,t]
\begin{center}
\includegraphics[scale=1]{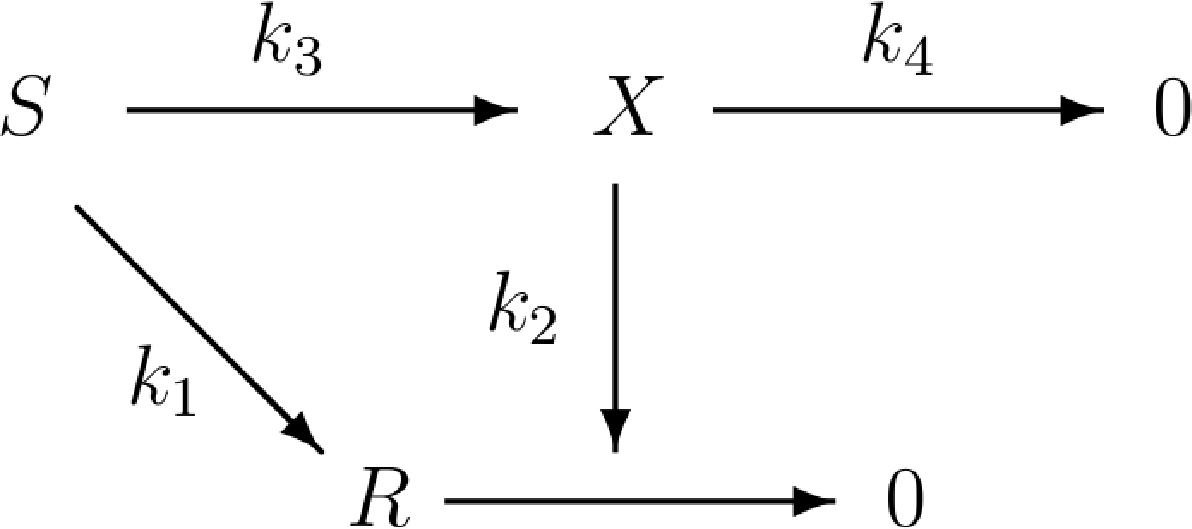}
\caption{``Sniffer'' network, from ~\protect{\cite{sniffers}}}
\label{sniffer_reactions}
\end{center}
\end{figure}

\newpage

\begin{figure}[h,t]
\begin{center}
\includegraphics[scale=0.8]{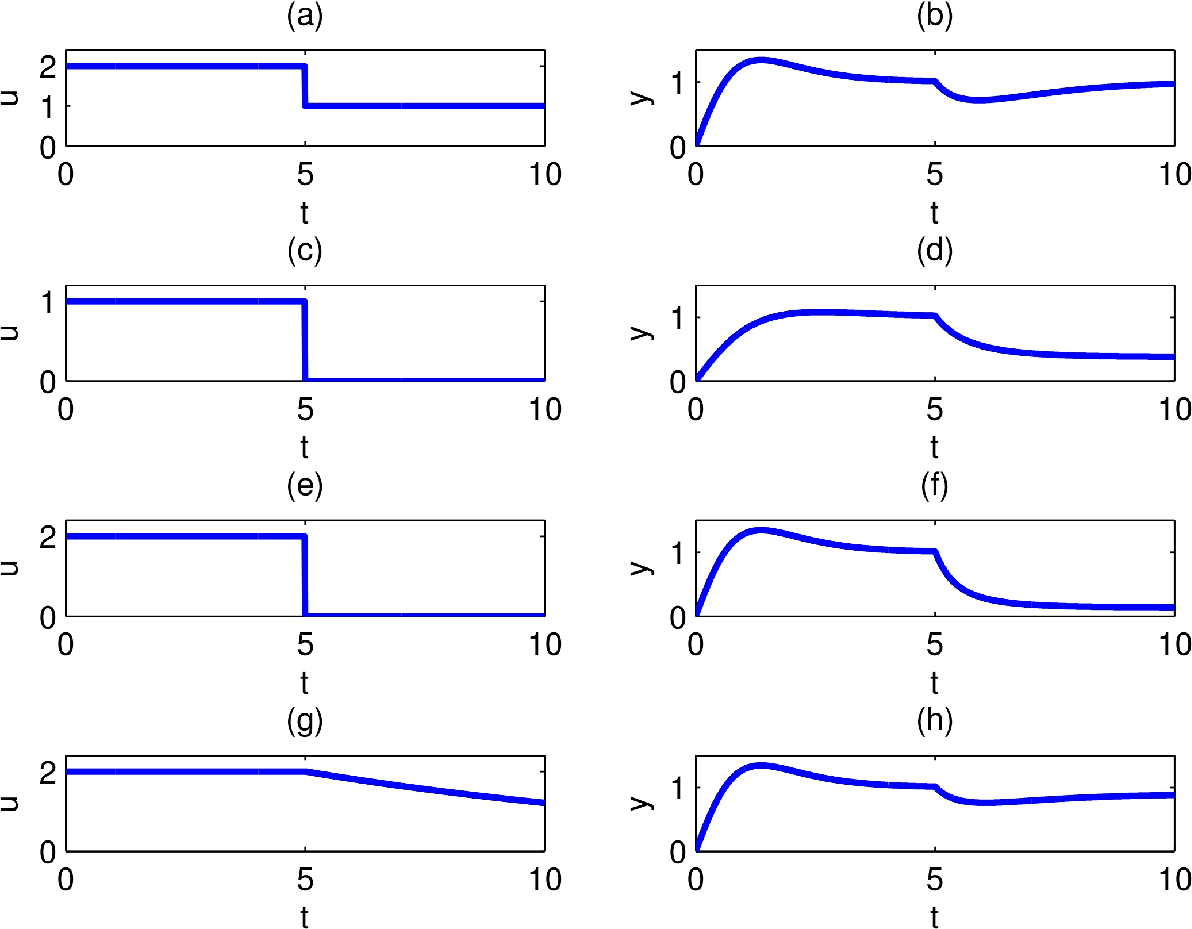}
\label{fig:4inputs}
\caption{%
Plots for example discussed in text.  In left column are inputs $u(t)$,
in right column are responses $y(t)$.  Initial conditions are $x(0)=y(0)=0$.
$u(t)$ is constant up to time $t=5$, switches to another constant value
at time $t=5$.
(a,b) $u$ switches to a nonzero value, $y$ adapts again.
(c,d) Pulse of magnitude $1$: asymptotic value of $y$ is 
$\approx 1/e \approx 0.37$ after pulse.
(e,f) Pulse of larger magnitude $2$: asymptotic value of $y$ is 
smaller: $\approx 1/e^2 \approx 0.14$ after pulse.
(g,h) Exponential decaying $u(t)=2e^{(t-5)/10}$ after time $t=5$
results in return to close to adapted value $y(\infty )\approx 0.9$
($u(t)$ plotted only on interval $[0,10]$ for ease of comparison).}
\end{center}
\end{figure}

\newpage

\begin{figure}[h,t]
\begin{center}
\includegraphics[scale=0.5]{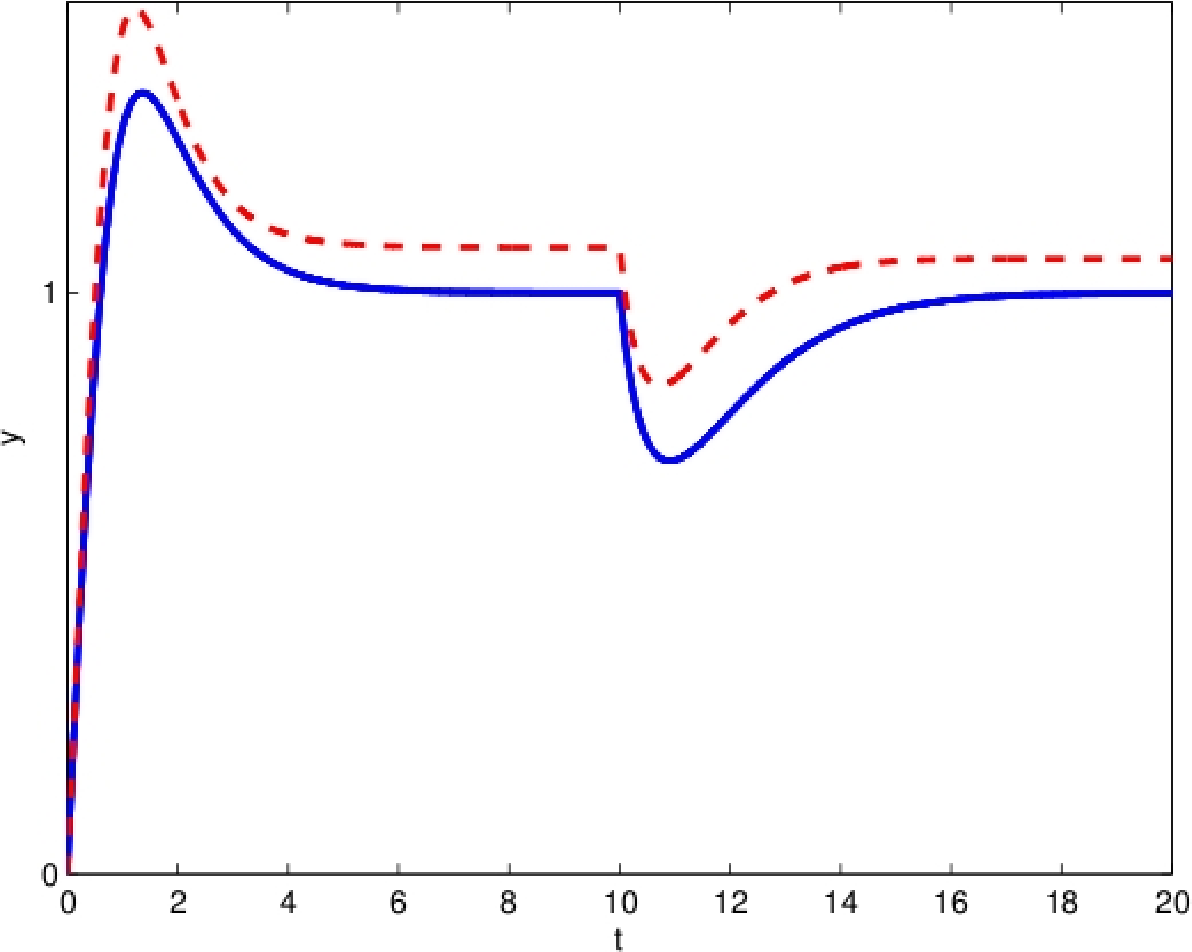}
\label{fig:ffpert}
\caption{Approximate adaptation: comparison of responses when right-hand side
  of equation is perturbed as discussed in text.  Input is as in
  Figure~\protect{\ref{fig:4inputs}}(a).  Solid line is for original system
  and dashed line for perturbed system.}
\end{center}
\end{figure}

\newpage

\begin{figure}[h,t]
\begin{center}
\includegraphics[scale=0.5]{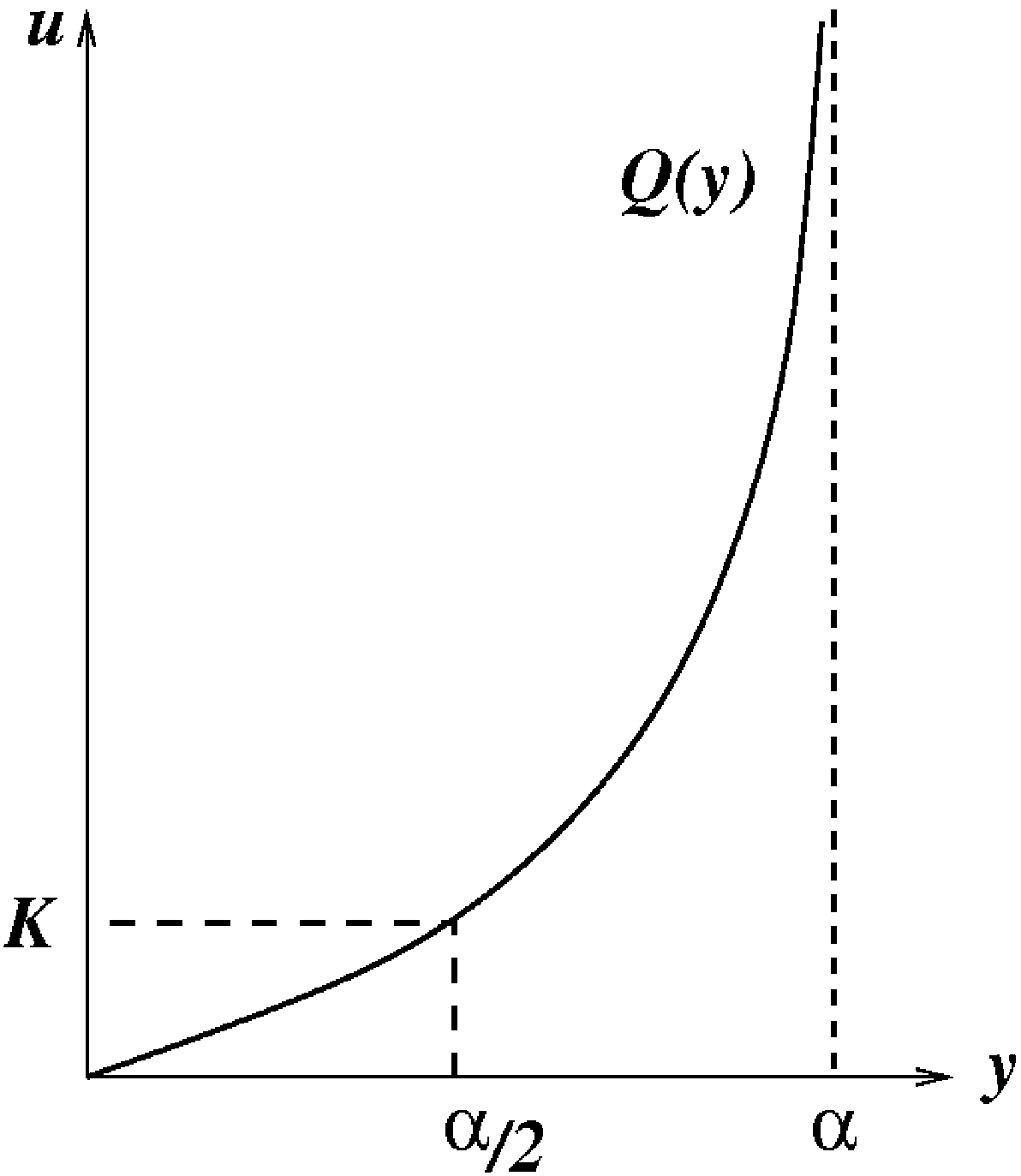}
\label{fig:Q}
\caption{The function $Q$}
\end{center}
\end{figure}

\newpage

\begin{figure}[h,t]
\begin{center}
\includegraphics[scale=0.7]{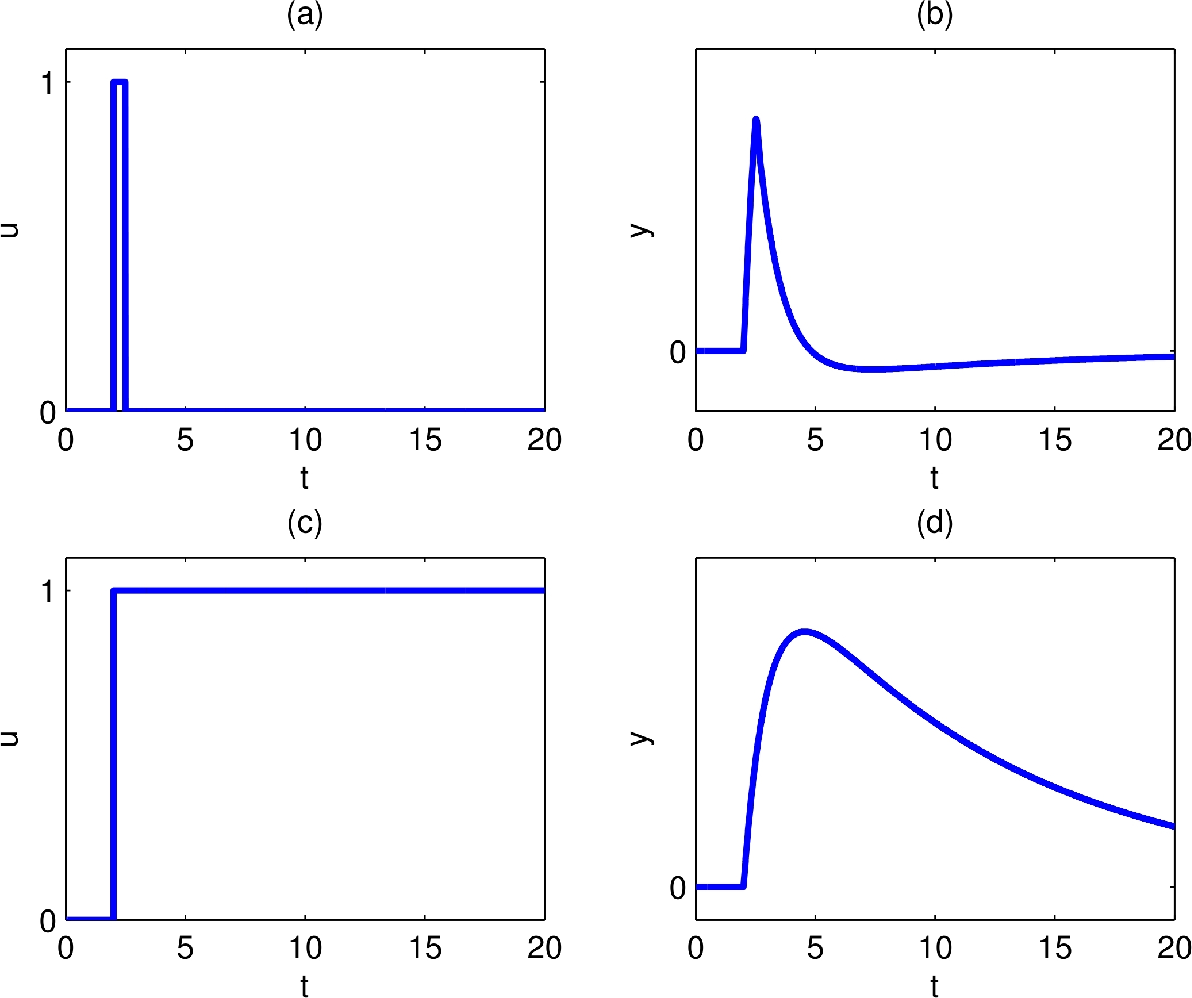}
\label{fig:fffb}
\caption{Responses to a pulse and to a step input, for a linear feedforward
  system discussed in the text.  (a) Pulse input.  (b) Response to the input
  in (a).  (c) Step input.  (d) Response to the input in (c).}
\end{center}
\end{figure}

\newpage

\begin{figure}[h,t]
\begin{center}
\includegraphics[scale=1]{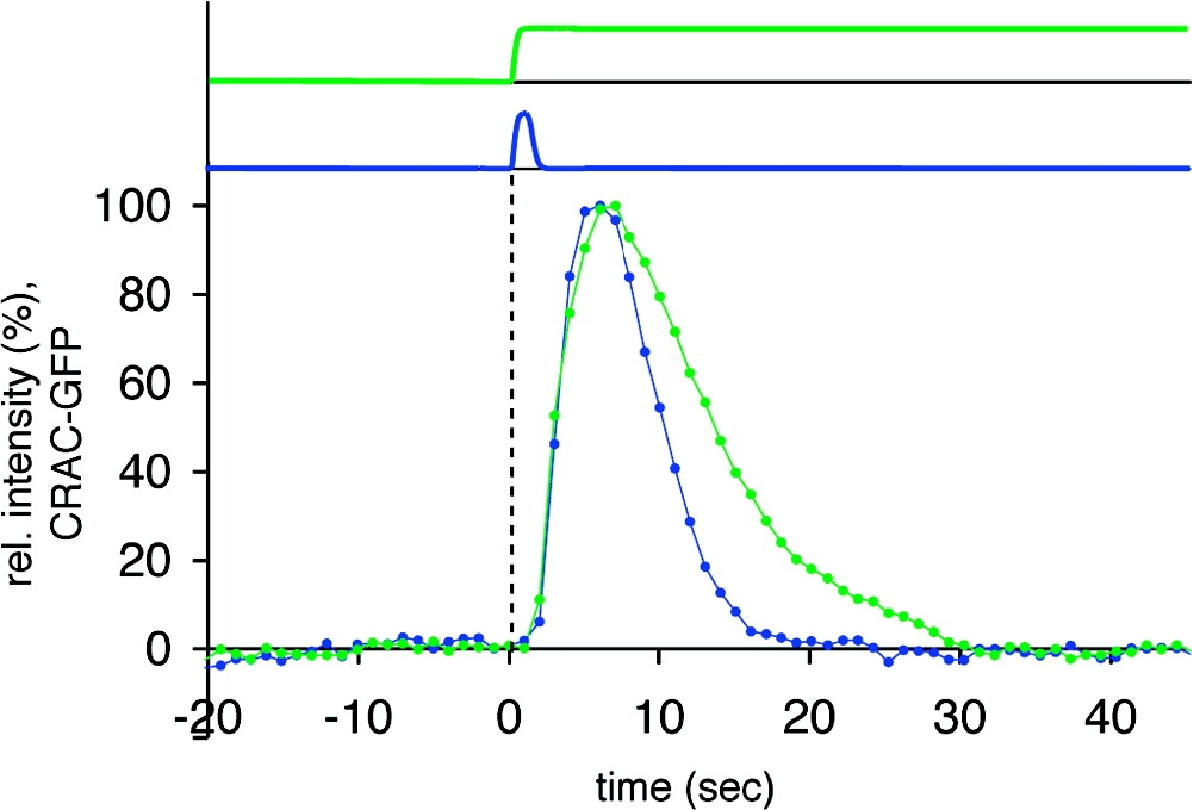}
\label{fig:bodenschatz}
\caption{Reproduced (microscopy inset removed) with permission from Figure 2
  in~\protect{\cite{bodenschatz}}.  Responses to step and pulse cAMP input
 in \emph{Dictyostelium}, as discussed in text.  The response to the pulse
 input settles faster than the response to the step.}
\end{center}
\end{figure}

\end{document}